\documentclass[a4paper]{article}
\usepackage{graphicx}

\usepackage{bm}

\begin{document}

\title{Stokes' drift: a rocking ratchet}
\author{I. Bena,  M. Copelli  and C. Van den Broeck \\ 
\ \\
Limburgs Universitair Centrum\\ B-3590 Diepenbeek, Belgium }
\date{\today}

\maketitle

\begin{abstract}
We derive the explicit analytic expression for the Stokes' drift in
one dimension in the presence of a dichotomic Markov forcing.  For
small amplitudes of the forcing, the drift is enhanced, but the
enhancement is reduced with increasing frequency of the forcing.  On
the other hand, a reduction of the drift or even a flux reversal can
be induced at larger amplitudes, while the flux is now found to be an
increasing function of the perturbation frequency.
\end{abstract}


\section{Introduction}
\label{sec:intro}

A longitudinal wave travelling through a fluid imparts a net drift
motion to the suspended particles - an effect known as {\it Stokes'
drift}.  The {\it classical} Stokes' drift~\cite{stokes} refers only
to the deterministic behavior, i.e. it does not account for the
stochastic fluctuations or perturbations in the system.  It has been
extensively studied in various practical contexts like, for instance,
the motion of tracers in meteorology and oceanography~\cite{meteo} and
that of the (doping) impurities in crystal growth~\cite{lee}.  A
simple intuitive explanation of classical Stokes' drift is that the
suspended particles spend a longer time in the regions of the
wave-train where the force due to the wave acts in the direction of
wave's propagation than the time spent in those regions where the
force acts in the opposite direction; therefore particles are driven
in the sense of wave's propagation.  When several linearly superposed
waves are present, the resulting drift velocity is simply the sum of
the contributions from each wave~\cite{hert}.

Recently, there has been some interest in {\it stochastic} Stokes'
drift~(\cite{hert,lythe,chris}, see also~\cite{landauer}).  It was
found that the diffusive motion of the suspended particles can
significantly alter the amplitude of the drift velocity.  But it is
the merit of~\cite{lythe} to open stochastic Stokes' drift to further
investigation by putting it in its natural context, namely {\it
ratchet-like Brownian motors} in which the thermal motion of small
(microscopic) particles is rectified by an asymmetric time-dependent
potential~\cite{ratchets}.  This scenario is referred to as the {\it
flashing ratchet}.  Paper~\cite{lythe} is limited to the case when the
motion of the overdamped suspended particles is diffusion-dominated,
i.e. when the deterministic forcing due to the wave is very weak.  An
approximate perturbative technique is developed for calculating the
drift velocity in arbitrary dimensions, which shows that the drift
depends on particles' diffusivity in magnitude and, in more than 1D,
also in direction (a very appealing feature, useful in particle
selection devices - see, for example, in~\cite{rratchet} Doering et 
all, 1994, and in~\cite{ratchets} M. Bier,
1996).  Papers \cite{chris} and, in a very different
context~\cite{landauer}, give an exact integral expression for Stokes'
drift of overdamped diffusive particles in 1D, for arbitrary wave
forms.  Using a Fokker-Planck equation technique, it finds that white
noise reduces Stokes' drift by comparison to its classical
(deterministic) value, the flux reversal being absent.

The purpose of this short communication is to show that the
characteristics of the 1D Stokes' drift may become more complex
(including flux reversal or enhancement of the classical drift) when
the particles are subjected to stochastic forcing that is non-white
(i.e. it has a finite correlation time).  We will focus on the case of
dichotomic Markov forcing, in which the particle is subjected to a
randomly alternating force, in addition to the force exerted by the
wave.  For this choice of coloured noise, explicit analytic results
can be obtained, while this is not the case for other types of
coloured noise, including the Ornstein-Uhlenbeck process.  Even though
the time-average of the dichotomic force is taken to be zero, its
effect in conjunction with the asymmetry of the wave's perturbation is
highly nontrivial.  As such the model realizes in the Stokes' drift
context another paradigm of Brownian motors, namely that of the
so-called {\it rocking ratchet}.  In this scenario, the systematic
motion acquired under influence of an alternating zero-average force
essentially revolves on the asymmetry of the nonlinear response in the
presence of a steady but asymmetric potential~\cite{rratchet}.  The
dichotomic Markov forcing presents another important advantage in that
the exact integral expression for Stokes' drift can be derived. 
Furthermore, in some cases of interest, the resulting integrals can be
evaluated explicitly.  Our analytic results are also complemented with
computer simulations, which are found to be in full agreement with the
theory.

\section{The model}
\label{sec:II}

The starting point is the following stochastic differential equation:

\begin{equation}
\dot{X}_t = f(X_t-vt) + \xi(t)\; ,
\label{eq1}
\end{equation}  
where $\xi(t)$ is a symmetric dichotomic noise (a two-step Markov
process)~\cite{vankampen}.  It can take only two values, $\pm A$ , with
equal probability, jumping between them with a probability $k$ per
unit time.  It has zero mean and its autocorrelation function
$\left<\xi (t)\,\xi(t')\right>\,=\,A^2\,\exp(-2\,k\,|t\,-\,t'|)$ shows
a finite correlation time $\tau _c\,= \,1/2k$.

With a suitable time scaling, eq.(\ref{eq1}) models the 1D overdamped
motion of a Brownian particle subjected to a dichotomic forcing. 
$X_t$ is the particle position at time $t$, while $f$ is the periodic
forcing due to the wave travelling at the speed $v$ with wavelength
$\lambda$:
\begin{equation}
f(y + \lambda) = f(y)\; , \forall y\; ,
\label{eq3}
\end{equation} 
with $f(y) < v\;,\forall y$ as the case of physical relevance.

The quantity of interest is the {\it drift velocity}, which represents
the mean velocity of the particles in the long time limit:
\begin{equation}
\vartheta =  \lim_{t\rightarrow\infty}
\frac{1}{t}\left<X_t\right>\; ,
\label{eq5}
\end{equation} 
where the mean is to be taken over  the realisations of the
dichotomic noise.  By defining a new
variable $Y_t=X_t-vt$ one can rewrite eq.(\ref{eq1}) as:
\begin{equation}
\dot{Y}_t = F(Y_t) + \xi(t)\; ,
\label{eq7}
\end{equation} 
with $F(y) \equiv f(y) - v$, and therefore:
\begin{eqnarray}
\vartheta - v & = & \lim_{t\rightarrow\infty} 
\frac{1}{t} \left<Y_t\right> \nonumber\\
& = & \lim_{t\rightarrow\infty} \int_{-\infty}^{+\infty}
dy\,F(y)\,[\rho_{+}(y,t) + \rho_{-}(y,t)]\; .
\label{eq9}
\end{eqnarray} 
Here $\rho_{\pm} (y,\,t)$ represent the probability densities for the
particle position $y$ and the value of the dichotomic noise equal to
$\pm A$, respectively.  Their  time evolution
is described by  \cite{vankampen}:
\begin{eqnarray}
\frac{\partial \rho_{+}(y,t)}{\partial t} & = & 
-\frac{\partial}{\partial y}[(F+A)\rho_{+}] 
-k(\rho_{+}-\rho_{-}), \nonumber\\
\frac{\partial \rho_{-}(y,t)}{\partial t} & = & 
-\frac{ \partial}{\partial y}[(F-A)\rho_{-}]  
-k(\rho_{-}-\rho_{+}).
\label{eq10}
\end{eqnarray} 
The first term in the r.h.s. of each of these equations corresponds to
the deterministic flow in phase space (between two jumps of the
dichotomic noise), while the other term corresponds to the jumps of
the dichotomic noise between $\pm A$.  Because of the periodicity
(\ref{eq3}) of $f(y)$ ($F(y)$), in order to evaluate (\ref{eq9}) we
only need to study the asymptotic behavior of the reduced probability
density:
\begin{equation}
P(y,t)=\sum_{n = -\infty}^{\infty} 
\left[
\rho_{+}(y + n\lambda,t) 
+\rho_{-}(y + n\lambda,t) 
\right].
\label{eq11}
\end{equation} 
From its definition, it is obvious that the reduced probability
density is subjected to periodic boundary conditions:
\begin{equation}
P(y+\lambda,t)=P(y,t)\; , \ \ \ \ \ \ \forall y,t,
\label{eq12}
\end{equation} 
while the normalization condition reads:
\begin{equation}
\int_{0}^{\lambda}P(y,t)\, dy=1.
\label{eq13}
\end{equation} 
Considering also the quantity
\begin{equation}
p(y,t) = \sum_{n = -\infty}^{+\infty} 
\left[
\rho_{+}(y + n\lambda,t)
-\rho_{-}(y + n\lambda,t) 
\right]\; ,
\label{eq14}
\end{equation} 
it follows from~(\ref{eq10}) that $P(y,t)$ and $p(y,t)$ obey a set
of coupled evolution equations:
\begin{eqnarray}
\frac{\partial P(y,t)}{\partial t} 
& = & - \frac{\partial}{\partial y}(FP)
      -A\frac{\partial}{\partial y}p , \nonumber\\
\frac{\partial p(y,t)}{\partial t} 
& = & -  \frac{\partial}{\partial y}(Fp)
      -A\frac{\partial}{\partial y} P-2kp.
\label{eq15}
\end{eqnarray} 
The stationary solution $P^{st}(y)$ can be easily
found~\cite{horsthemke}:
\begin{equation}
P^{st}(y)=\frac{1}{Z}\int_{y}^{y+\lambda}
\frac{F'(y')+2k}{F^{2}(y')-A^2}
\exp[-\phi(y')+\phi(y)]\, dy',
\label{eq23}
\end{equation}
which is valid as long as $F^2(y)-A^2\neq 0$, $\forall y$.  Here
$\phi(y)$ is defined by:
\begin{equation}
\phi(y)=
-2\int_{0}^{y}\frac{F(y')[F'(y')+k]}{F^2(y')-A^2}dy'.
\label{eq20}
\end{equation} 
and $Z$ is the normalization factor according to eq.(\ref{eq13}). 
Therefore the drift velocity can be finally expressed as:
\begin{equation}
\vartheta=\int_{0}^{\lambda}f(y)P^{st}(y)\,dy\; .
\label{eq25}
\end{equation} 
These results are valid for an arbitrary wave form, provided that the
obvious necessary differentiability and integrability conditions are
fulfilled.  Note that the drift depends nonlinearly on $F$, so that
the contributions of different waves are not additive.

\section{The piecewise linear wave}
\label{sec:III}

An explicit  analytical expression for $P^{st}(y)$ can be  
obtained  in the case 
of a piecewise linear wave forcing:
\begin{equation}
F(y)=
\left\{
\begin{array}{ll}
-(1-b)v, 
& y \in [0,\lambda/2-2\varepsilon[ \; \mbox{mod} \; \lambda \\
- [(1-b)v
+bv(y-\lambda/2+2\varepsilon)/\varepsilon], 
& y \in [\lambda/2-2\varepsilon,\lambda/2[ \; \mbox{mod} \;\lambda \\
-  (1+b)v, 
& y \in  [\lambda/2,\lambda-2\varepsilon[  \; \mbox{mod} \; \lambda \\
- [(1+b)v 
-bv(y-\lambda+2\varepsilon)/\varepsilon], 
& y \in [\lambda-2\varepsilon,\lambda[   \; \mbox{mod} \; \lambda ,
\end{array}
\right.
\label{eq26}
\end{equation}
with $0<b<1$.  A square wave corresponds to $\varepsilon\rightarrow 0$,
while $\varepsilon =\lambda/4$ represents the triangular-like wave.

\begin{figure}[tb]
\begin{center}
\includegraphics[width=0.9\textwidth]{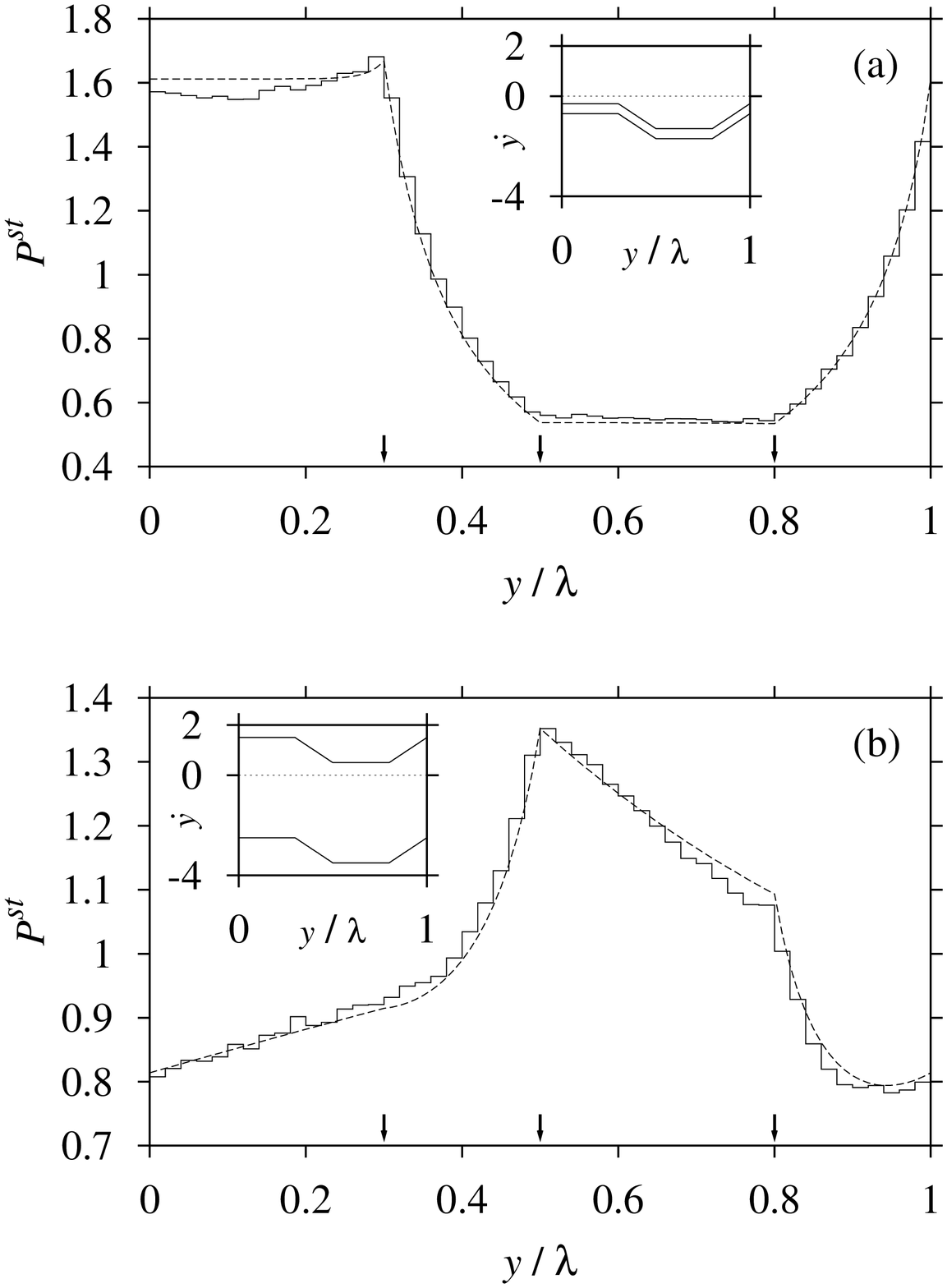}
\caption{\label{fig:Py} Stationary probability density for the
piecewise linear wave, eq.~(\ref{eq26}), with $v=1$, $b=0.5$, $\lambda
= 10$ and $\varepsilon = 1$ (the arrows indicate the boundary points
of the piecewise linear regions).  Dashed lines represent the
theoretical results, while solid lines correspond to simulations of
$10^{6}$ particles evolving for 200 time units.  The insets show the
two realisations of $\dot{y}$.  Fig.~\ref{fig:Py}(a): $A=0.2$, $k=1$. 
Fig.~\ref{fig:Py}(b): $A=2$, $k=0.1$.}
\end{center}
\end{figure}

The final expressions for $P^{st}(y)$ are rather lengthy and will not
be reproduced here.  But we performed numerical simulations, which
show very good agreement with theoretical results, cf. 
fig.~(\ref{fig:Py}).  Note that these results are valid for
$0<A<(1-b)v$ or for $A>(1+b)v$.

The analytic results for the square-like wave, corresponding to the
limit $\varepsilon \rightarrow 0$, are of more interest.  They read as
follows:
\begin{equation}
P^{st}(y) = Z^{-1}\left\{ \frac{1-e^{-U-V}}{(1-b)v}
+\frac{e^{2Uy/\lambda}}{(1-b^2)v} \left[ e^{-U}-e^{-U-V} \right]
\frac{2A^2b}{(1-b)^2v^2-A^2}\right\}\; ,
\label{eq28}
\end{equation} 
for $y\in [0,\lambda/2[\;$, and 

\begin{equation}
P^{st}(y) = Z^{-1}\left\{ \frac{1-e^{-U-V}}{(1+b)v}
-\frac{e^{V(2y/\lambda-1)}}{(1-b^2)v} \left[ e^{-V}-e^{-U-V} \right]
\frac{2A^2b}{(1+b)^2v^2-A^2}\right\}\; ,
\label{eq29}
\end{equation}
for $y\in [\lambda/2,\lambda[\;$, where we introduced the
dimensionless variables:
\begin{eqnarray}
U & \equiv & 
\frac{{\alpha}(1-b)}{2[(1-b)^2{\beta}^2-1]}\; ,\nonumber\\ 
V   & \equiv & 
\frac{{\alpha}(1+b)}{2[(1+b)^2{\beta}^2-1]}\; , \nonumber \\ 
{\alpha} & \equiv & \frac{2kv\lambda}{A^2},\ \ \ \ \ 
{\beta}\equiv \frac{v}{A} 
\label{eq30}
\end{eqnarray} 

The following expression for the drift velocity is obtained:
\begin{equation}
\frac{\vartheta}{\vartheta_{cl}} = 
\frac{1-e^{{U}+{V}}-
\frac{4}{(1-b^2){\alpha}}
(1+e^{U+V}-e^{U}-e^{V})}
{1-e^{U+V}-\frac{4b^2}{(1-b^2)
{\alpha}}(1+e^{U+V}-e^{U}
-e^{V})} ,
\label{eq31}
\end{equation} 
where $\vartheta_{cl}=b^{2}v$ is the classical value of the Stokes'
drift velocity.  We mention some limiting cases of interest.  First,
the limit $A\rightarrow 0$ or $k\rightarrow \infty$ (i.e.
$\beta\rightarrow\infty$ and/or $\alpha\rightarrow\infty$) leads to
the classical value of the Stokes' velocity
${\vartheta}/{\vartheta_{cl}}=1$.  Second, the drift velocity derived
in~\cite{chris} is recovered in the white noise limit $A\rightarrow
\infty$, $k\rightarrow\infty$, with $A^2/2k=D$ finite ($\beta \rightarrow 0$). 
Finally, the quenched-noise limit $k\rightarrow 0$ (${\alpha} \rightarrow 0$)
(i.e. half of the particles, chosen at random, are subjected to a constant
external forcing $+A$, while the other half to an external forcing
$-A$) results in the following mean velocity:
\begin{equation}
\frac{\vartheta_{qn}}{\vartheta_{cl}}=\frac{{\beta}^2}{{\beta}^2-1}.
\label{eq36}
\end{equation} 
This expression reveals the possibility of a {\it flux reversal}: the
drift velocity is negative for ${\beta} <1$, i.e. for sufficiently
large amplitude.

\begin{figure}
\begin{center}
\includegraphics[width=0.9\textwidth]{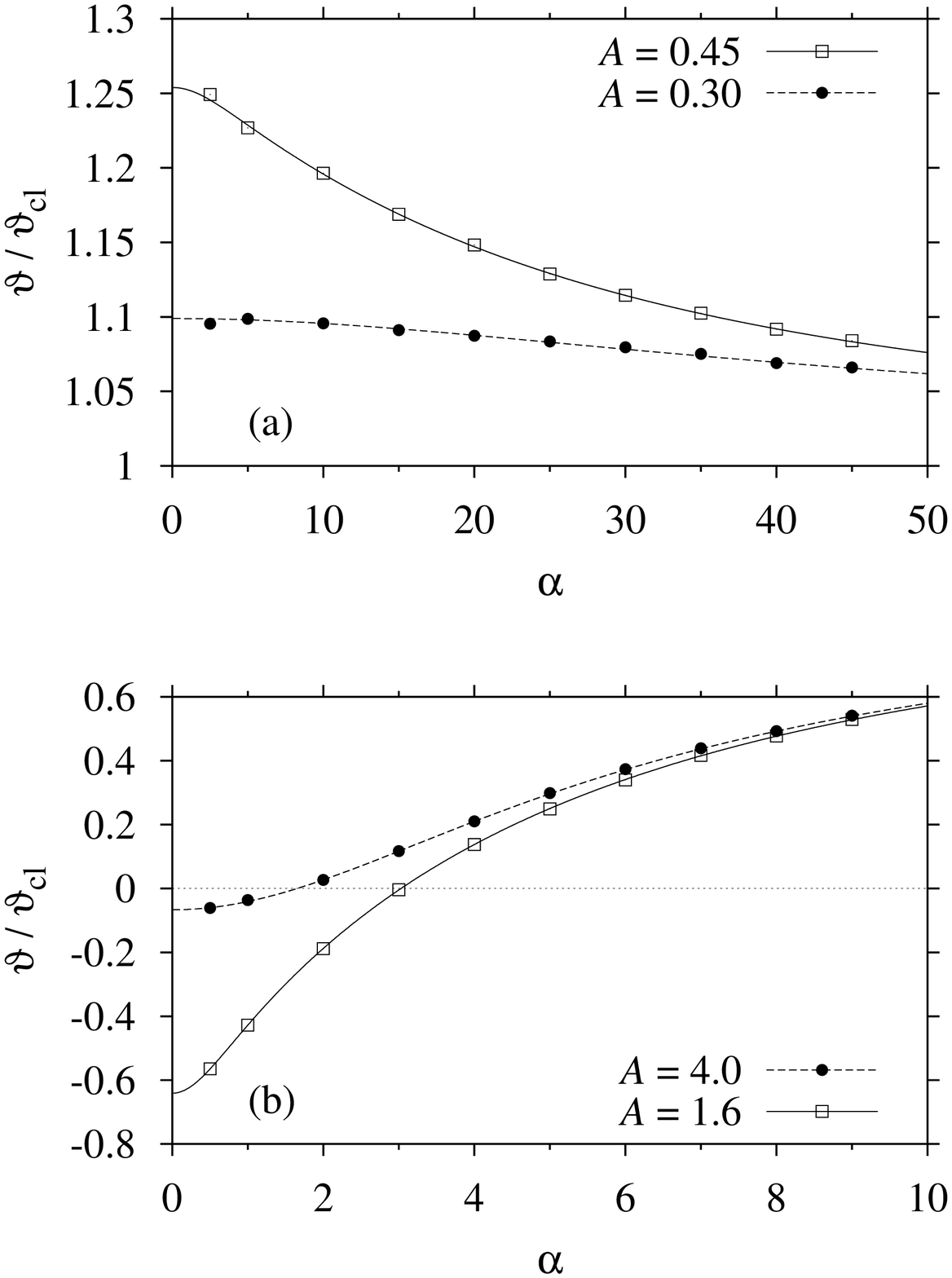}
\caption{\label{fig:drift}Drift velocity according to eq.~(\ref{eq5})
for a square wave with the same values of the parameters as in
Fig.~\ref{fig:Py}, except here $\varepsilon=0$.  Symbols represent
simulations with $10^{3}$ particles measured after $5\times 10^{4}$
time units.  The lines represent the theoretical value,
eq.~(\ref{eq31}).  Fig.~\ref{fig:drift}(a): drift enhancement. 
Fig.~\ref{fig:drift}(b): flux reversal.}
\end{center}
\end{figure}

Turning to the general case, we have represented in
Fig.~\ref{fig:drift} the results of eq.~(\ref{eq31}) together with
simulations for the square wave for two representative sets of
parameter values.  For small values of the forcing amplitude $A$ one
finds that the drift is enhanced ($\vartheta/\vartheta_{cl}>1$), but
the enhancement is reduced upon increasing the frequency of the
perturbation.  In fact one can prove that enhancement occurs whenever
$A<(1-b)v$, by noting that $U,V>0$ and making use of the inequality
$e^{x}\geq 1+x$.  On the other hand, a reduction of the drift or even
a flux reversal ($\vartheta < 0$) can be induced at larger amplitudes
(for small $k$, according to eq.~(\ref{eq36})), while the flux is now
found to be an increasing function of the perturbation frequency.

\section{Concluding remarks}
\label{sec:IV}

In conclusion, the rocking ratchet version for Stokes' drift reveals
two new features, which are to be attributed to the colour (finite
correlation time) of the applied force: the drift is enhanced by small
amplitude perturbations, while a flux reversal occurs at sufficiently
large amplitude and large correlation time.  While our calculations
are limited to the case of dichotomic Markov noise, we expect that
similar results hold for other coloured noises.

\begin{figure}
\begin{center}
\includegraphics[width=0.65\textwidth]{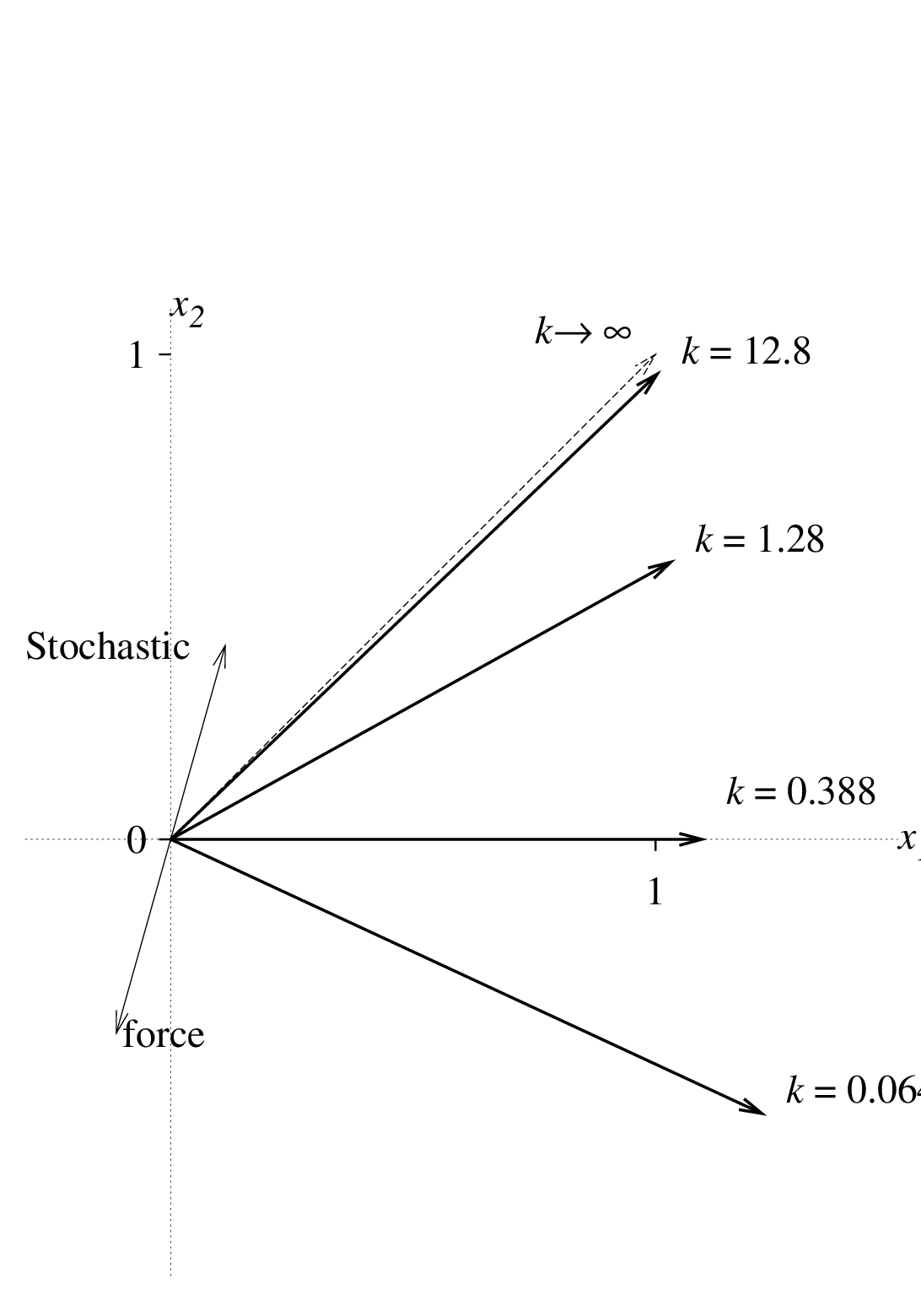}
\caption{\label{fig:2D}The resulting drift velocity (in units
$\vartheta_{cl}$; thick arrows) for two identical square waves (with
the same parameters as in Fig.~\ref{fig:Py}) propagating respectively
along the $+\,x_{1}$
and $+\,x_{2}$ axes, in the presence of a stochastic forcing with 
$A_{1} =~0.45$, $A_{2} = 1.6$ (applied along the direction of the thin arrows),
for different transition rates $k$. 
The dashed arrow corresponds to the classical limit $k\to\infty$.}
\end{center}
\end{figure}

In~\cite{lythe} it was reported that, in two dimensions, both the
direction and amplitude of the Stokes' drift can change with the
intensity of the Brownian motion when several waves are present.  A
similar phenomenon will appear here when one considers, for example,
two waves propagating in orthogonal directions $x_{1}$ and $x_{2}$.  A
dichotomic forcing $\bm{\xi}$ will induce different drifts along these
directions depending on its projections $\xi_{1} = \pm A_{1}$ and
$\xi_{2} = \pm A_{2}$, as well as on its transition rate $k$.  In 
particular, by
varying $k$  one is able to induce a significant change in the
resulting drift direction, as can be seen in fig.~\ref{fig:2D} for the
simple example of identical square waves.

\paragraph{Acknowledgements}
This work was supported by the Program on Inter-University Attraction
Poles of the Belgian Government and the F.W.O. Vlaanderen (CVdB).


\begin{thebibliography}{99}

\bibitem{stokes} G. G. Stokes, Trans. Camb. Philos. Soc. {\bf 8}, 441 
(1847);
O. M. Phillips, {\it The Dynamics
of the Upper Ocean}, Cambridge University Press, Cambridge, 1977;
J. Lightwill, {\it Waves in Fluids}, 
Cambridge University Press, Cambridge, 1978.


\bibitem{meteo} A few recent papers from a rich bibliography on the 
subject:\\
A. M. Bratseth, Tellus {\bf 50A}, 451 (1998);
S. M. Cox, Fluid Dyn. Res. {\bf 19}, 149 (1997);
H. C. Graber, B. K. Haus, R. D. Chapman and  L. K. Shay, 
J. Geophys. Res. {\bf 102}, 18749 (1997) (experimental);
J. C. McWilliams, P. P. Sullivan and 
Chin-Hoh Moeng, J. Fluid Mech. {\bf 334}, 1 (1997);
M. Nordsveen and A. F. Bertelsen, 
Int. J.   Multiphase Flow {\bf 23}, 503 (1997);
V. Polonichko, J. Geophys. Res. {\bf 102}, 15773 (1997);
S. A. Thorpe, J. Phys. Oceanogr. {\bf 27}, 62 and 2072 (1997).

\bibitem{lee} See e.g. a recent paper of C. P. Lee, Phys. of Fluids 
{\bf 10},
2765 (1998) and references therein.

\bibitem{hert} K. Herterich and K. Hasselmann, 
J. Phys. Oceanogr. {\bf 12}, 704 (1982);
O. N. Mesquita, S. Kane and J. P. Gollub, Phys. Rev. A {\bf 45}, 3700 
(1992).

\bibitem{lythe} K. M. Janson and G. D. Lythe, 
Phys. Rev. Lett. {\bf 81}, 3136 (1998); 
ibid., J. Stat. Phys. {\bf 90}, 227 (1998).

\bibitem{chris} C. Van den Broeck, Europhys. Lett. {\bf 46}, 1 (1999).

\bibitem{landauer} R. Landauer and M. B\"uttiker, Physica Scripta {\bf
T9}, 155 (1985).

\bibitem{ratchets} M. O. Magnasco, Phys.  Rev.  Lett.  {\bf 71}, 1477
(1993); M. Bier, Phys.  Lett.  A {\bf 211}, 12 (1996); P. H\"{a}nggi
and R. Bartussek, in {\it Lect.  Notes in Phys.}, vol.  {\bf 476}, ed. 
by J. Parisi et al., Springer, Berlin, 1996; R. D. Astumian, Science
{\bf 276}, 917 (1997); F. Julicher, A. Ajdari and J. Prost, Rev.  Mod. 
Phys.  {\bf 69}, 1269 (1997).

\bibitem{rratchet}
C.R. Doering, W. Horsthemke and J. Riordan,Phys. Rev. Lett. {\bf 72}, 2984
(1994); K.W. Kehr, K. Mussawisade, T. Wichmann and W. Dieterich, Phys. Rev. E
{\bf 56}, R2351  (1997).

\bibitem{vankampen} N. G. Van Kampen, {\it Stochastic Processes in Physics and 
Chemistry}, North Holland, Amsterdam, 1981.

\bibitem{horsthemke} W. Horsthemke and R. Lefever {\it Noise-Induced
Transitions : Theory and Applications in Physics, Chemistry and
Biology}, Springer Verlag, 1984.





\end{thebibliography}
\end{document}